# Seer: Empowering Software Defined Networking with Data Analytics

Kyriakos Sideris, Reza Nejabati, Dimitra Simeonidou
High Performance Networks Group
University of Bristol, UK
kyriakos.sideris@bristol.ac.uk

*Abstract–Network complexity is increasing, making network control and orchestration a challenging task. The proliferation of network information and tools for data analytics can provide an important insight into resource provisioning and optimisation. The network knowledge incorporated in software defined networking can facilitate the knowledge driven control, leveraging the network programmability. We present Seer: a flexible, highly configurable data analytics platform for network intelligence based on software defined networking and big data principles. Seer combines a computational engine with a distributed messaging system to provide a scalable, fault tolerant and real-time platform for knowledge extraction. Our first prototype uses Apache Spark for streaming analytics and open network operating system (ONOS) controller to program a network in real-time. The first application we developed aims to predict the mobility pattern of mobile devices inside a smart city environment.*

## I. Introduction

A big growth in networks has been observed in recent years, ranging from end-user mobile data networks through to machine-to-machine communications. The number of devices connected to IP networks will continue to increase up to three times the global population in 2020, while the global Internet traffic will be equivalent to 95 times the volume of the entire global Internet in 2005 [1]. As the complexity and interdependence of networks increase, management will be a progressively challenging task.

The importance of data analytics has increased due to the fact that data can fuel opportunities across many disciplines. For instance, businesses can deliver personalised services in social media and e-commerce, and concurrently researchers can understand and investigate the secrets of the human genome. Data analytics inherited a plethora of statistics, machine learning and data mining algorithms, only to combine with scalable and fault tolerant computer science methodologies and finally deliver big data tools. Network orchestrators and cloud providers are enabled to optimise their services by using big data.

On the other hand, the software defined networking (SDN) paradigm facilitates network evolution and innovation by simplifying network hierarchy and separating the control from data planes. The extracted network knowledge can drive intelligent SDN control by providing a clearer view of the network dynamics. The importance of network analytics increases as SDN is further established in the networking market [2].

Although various frameworks have been proposed to resolve network optimisation problems using big data analytics, they do not propose an architecture that can work as common ground for algorithm development. Moreover, reliability is becoming increasingly important as network elements become main components of the control plane. Therefore, any proposed platform should meet the challenges of high availability and scalability whilst being adaptive enough to carry out a diverse range of tasks. In addition, it should also run on commodity hardware in order to reduce the capital expenditure.

In this paper we introduce Seer, a flexible, highly configurable data analytics platform for network intelligence based on SDN and big data principles. Seer uses active or passive network elements, which are either SDN enabled or not, to extract information related to the network state. Various sources of information are combined in order to generate knowledge, which in turn is used to leverage SDN, either by optimising the network resource allocation or by predicting the next state and proactively orchestrating resources. The extracted knowledge is also presented to network applications in the higher layers to assist their management processes.

Our platform provides the benefits of big data architectures on network data management and data analytics [3] to the SDN network paradigm in order to leverage network programmability, ease of management, optimisation of network resources and eminent failure recovery [4]. The main improvement of Seer lies in the fact that the platform has been built on mature open-source tools, using design methodologies employed in industry. All elements of the proposed platform run on commodity hardware and can adapt to any network analytics related task.

The applications running on Seer can vary from core to edge networks, targeting storage, computing or orchestration applications. The developer defines the type of extracted information from the network, the format that will be used to disseminate information and the model under which this information will be combined in order to generate the necessary knowledge. In addition, the implementation of new algorithms targeting Seer's controllers and computation engines is straightforward, as it is based on well-established

The work presented in this paper has been supported by EPSRC Knowledge Centric Networking (KCN) project (EP/L026120/1) and EPSRC Towards Ultimate Convergence of All Networks (TOUCAN) project (EP/L020009/1).

technologies. A mobility management application was developed on this platform in order to highlight the benefit of using Seer for dynamic resource optimisation when used in a chain of services in a smart city environment.

The remainder of the paper is organised as follows. Section II provides the background on this work, whilst related work is presented in Section III. The main components of the platform are presented in Section IV as well as our first prototype. Section V describes the mobility management use case, as the first application for evaluation, followed by Section VI which concludes the discussion.

## II. BACKGROUND

### A. Major Concepts

The Seer architecture was inspired by three major concepts: *i*) knowledge extraction based on techniques used for processing of large datasets; *ii*) knowledge exchange to fulfil network management goals; and finally *iii*) knowledge utilisation to drive programmable networks.

*1) Knowledge Centric Networking:*

In the knowledge centric networking (KCN) paradigm, the terms 'data', 'information' and 'knowledge' have different semantics and their use is not interchangeable. The term 'data' is related to raw facts retrieved from the network, whereas 'information', or *D1 knowlet*, describes processed, aggregated and filtered data that represent snapshots of the system's state. 'Knowledge', or *D2 knowlet*, represents the outcome of cognitive and analytical processes based on given sets of information. Using the example of Wi-Fi mobility, the probe request, authentication and association packets exchange represent raw facts that describe the roaming procedure of a Wi-Fi node. The aggregation of the above raw information comprises a D1 knowlet that describes a snapshot of the network state with respect to a Wi-Fi node. The combination of repetitive time snapshots related to a certain Wi-Fi node illustrate the mobility pattern of the node, which forms a D2 knowlet, and can be used to accurately predict future directions of movement.

The extracted knowledge can be either exploited within a single administrative domain (AD) or shared between different systems. Three styles of knowledge exchange are formed depending on the relation between knowledge producer and consumer: Knowledge as a tool (KaaT) describes the knowledge exchange that enables intelligent service or network control within a single AD, whilst knowledge as a service (KaaS) represents the knowledge transfer across different domains as a provided service based on service level agreements (SLAs) in order to generate business opportunities. In the KaaT style, the knowledge producer (KP) and knowledge consumer (KC) belong to the same AD, whilst in KaaS they belong in different ADs. The third knowledge exchange style is knowledge as a cloud (KaaC) which embodies open and flexible sharing that is not based on SLAs. In the previously described Wi-Fi case, the exploitation of the mobility pattern knowledge with the aim of reducing handover time delay within a Wi-Fi domain constitutes KaaT. If the mobility pattern knowledge is shared with a content delivery network under an SLA, it becomes KaaS and receives characteristics of inter-service communication. If the user is providing its mobility habits voluntarily to solve the road traffic congestion of the city is living in, then the mobility pattern knowledge becomes KaaC.

It is obvious that knowledge carries more value when it is presented to a relevant recipient. The KCN architecture advocates a distributed knowledge ecosystem where knowledge is harvested, clustered and presented to the right location. In order to separate the area of concern and ease network control, the KCN architecture emphasises on knowledge locality. Local SDN agents can collect and process local knowlets before reporting them to a central controller. Moreover, they can enforce local decisions without the global network view in cases where the AD holds a more relaxed policy [5], [6]. The locality of knowledge is particularly important in mobile edge computing scenarios [7].

*2) Software Defined Networking*

Software Defined Networking (SDN) is a very promising networking paradigm that aims to decouple the control logic from the forwarding system by promoting a logically centralised entity, also known as the network controller or network operating system[4]. The separation between the control plane, the logic that enforces network policies and take forwarding decisions, and the data plane, the underlying switching infrastructure, introduces a flexible way of monitoring, managing and evolving a network. SDN essentially makes network architecture simpler and the control plane more adaptive to changes. The resulting network programmability can be leveraged by enhancing the information provided to the control layer. Hence, network data extraction and knowledge dissemination is hence essential in projecting the correct network state and closing the control loop. Moreover, changes in the infrastructure can be foreseen and accommodated in a proactive and organised manner. The most notable software network controllers are OpenDaylight [8] and the open network operating system (ONOS) [9].

*3) Big Data Analytics*

Data analytics describe the science that examines raw data and aims to draw conclusions about the information that the data are carrying. Many industries realised the potential value of massive raw datasets, which they tried to unlock by using data analytics to drive decision making, a process often described as evidence-based decision making. The proliferation of data sources increased the volume of raw data as well as the diversity of their characteristics. The above change is expressed by the term "big data", which became widespread in 2011 [3]. Big data are popularly described by the five 'Vs': (*i*) volume; referring to the size of data set, (*ii*) variety; because of data's heterogeneous nature, (*iii*) velocity; describing the high rate of data production, (*iv*) value; indicating data usefulness, and finally (*v*) veracity; relating to the quality of data [10]. The data processing comprises of two main tasks: firstly, data management, which is formed of acquisition followed by extraction, cleaning and annotation and finishing with aggregation and representation. Secondly, data analytics, which usually follows data management, consists of modelling and analysis as well as interpretation [3][11]. Machine learning (ML) is the main vehicle of data analytics for prediction and optimisation, and is divided into three main categories: supervised, unsupervised, and reinforcement learning [12]. Systems that receive a high volume of data and require responsive behaviour rely on streaming, real-time, continuous, on-line analytics because of space and latency constrains.

*B. Motivation & Problem Statement*

In order to achieve high adaptivity in network control, a potential solution should address the following two requirements: *i)* accurate, punctual and reliable knowledge extraction and dissemination; and *ii)* knowledge driven network control. The next paragraphs elaborate on the key concepts of KCN and highlight the necessity of a platform to deliver the required knowledge to the control plane.

*1) Knowledge extraction & dissemination:*

Modern networks are considered complex systems with a high number of parameters and stakeholders. Network control and orchestration is tasked with taking actions on various parameters including bandwidth allocation, link utilisations, latency, jitter and energy consumption over a composition of heterogeneous networks types (i.e. optical, electrical, wireless). As the network state evolves with time, many parameters change, making the administration of fast changing systems, such as edge networks an extremely complex task. The more information we have from the network, the better the controller can adapt to any challenges presented. Moreover, the quantity and quality of information extracted is related to better representation of network's state and better prediction of the next state. The composability of the extracted information with other sources is also required in order to provide a clearer insight of the network. Therefore, the information should have commutative and associative properties, in space and time. As information from different sources is combined, the confidence interval of the resulting information decreases and stemming information matures to become knowledge. Well studied data mining and data analytics techniques can be used in the process of data filtering and aggregation as well as in the analysis and modelling of related information in order to derive network knowledge. Tools created for big data [13]-[16], can now facilitate the development of online algorithms for network analytics gaining in performance, scalability, manageability and debug-ability [17], [18]. We aspire Seer to meet this challenge in an open platform. Many messaging patterns have been proposed in the literature for service-oriented architecture including message type pattern, message channel pattern, message routing pattern etc. [19], [20]. The selection of a pattern depends on required characteristics like scalability, performance, availability, security etc.

*2) Knowledge driver control*

In order to optimise the resource allocation and avoid over-provisioning and violation of agreements, the control plane and cloud orchestrators can employ optimisation techniques based on multiple knowledge topics. Researchers have tried to tackle network related challenges by applying algorithmic approaches for P complexity class problems and heuristics for NP-complete problems on information extracted from the network [21]. Most of the literature emphasises on utilising internal network and cloud knowledge in order to achieve optimum performance. The separation between the control and forwarding planes, offered by SDN, enabled more flexible means of controlling the network so the control plane can consolidate independent sources of knowledge, in a logical centralised point, to target optimal performance. External information provided from other stakeholders can be factored in the network control and cloud orchestration process for the mutual benefit of all players. A content delivery network provider that exposes the most popular contents and their location can help the network provider to optimise network utilisation and enable a gain in quality-of-service (QoS) at cheaper rates. The orchestration of *X*-as-a-Service (XaaS) components like cloud computing [22], cache [23] and load balancing [24], [25], etc., can be linked to knowledge providing services in order to make network management easier.

III. RELATED WORK

The use of machine learning techniques to solve networking challenges has been extensively studied. The network optimisation aspects of the traditional network management, like traffic classification [26] and energy efficient resource management [27], as well as the dynamic resource management [28], have been thoroughly investigated. A traffic engineering framework with machine learning was also studied in SDN [29]. In addition, researchers have extensively used data mining techniques for security problems such as intrusion detection systems (IDS) [30]. The techniques used included supervised learning, (neural networks, decision trees, support vector machines, etc.), unsupervised learning (K-means clustering, principle component analysis etc.) and reinforcement learning (Markov decision processes etc.) [26], [27], [29]. Overall, the research community highlighted that a broad range of networking challenges can be met by employing a wide range of machine learning and data mining techniques.

Despite the benefits offered by algorithmic development, the community has not adapted a common platform to deliver research innovation to the real world. Most algorithms face scalability issues after leaving simulation or laboratory environments. In the case of operational critical algorithms, fault tolerant design should provide the stability of a production-ready subsystem. Moreover, most of the development has assumed offline execution, or execution on a big batch of data. Furthermore, there is little reference on externally assisted decision making and knowledge exchange to target optimal solution. The use of big data tools and techniques can solve the above shortcoming by bringing the inherited scalability, high availability, distributed and online processing to SDN. Since the quality of a platform depends on the properties of its constituents, a proposed design should combine elements that satisfy all the above properties.

IV. SEER

Seer is a platform that employs concepts related to big data, for data mining and data analytics to transform, analyse and model the information that has been extracted from the network with a goal to create a deep understanding of the network state. The inferred knowledge of the network is combined with knowledge provided from a third party in order to predict or optimise different aspects of the system by driving the control plane of software defined networks, storage, compute and orchestration. The platform is modular and is constructed by five main elements: *i*) data or information extraction, *ii*) data distribution, *iii*) data analytics, *iv*) knowledge dissemination, *v*) knowledge driven control. The basic components are also presented in Figure 1.

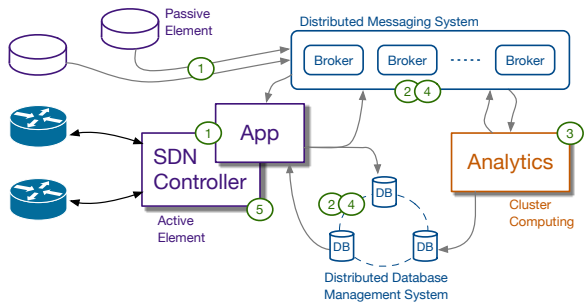

Figure 1. Basic components of the Seer platform. (1) data or information extraction, (2) data distribution, (3) data analytics, (4) knowledge dissemination, (5) knowledge driven control

### A. Modules and Functionality

The data or information extraction is responsible for acquiring network resources control data. The extraction mechanism can optionally apply basic format transformation, filtering or aggregation, depending on capabilities, before sending data for analysis. The mechanism can be either an active network element, that acts as monitor and actuator, or a passive, monitoring only, component. Examples of active elements include traffic engineering devices like packet forwarding devices, optical wavelength switches, traffic inspectors, traffic filter, etc. On the other hand, passive elements include Wi-Fi sniffers, optical power monitors, computing resources monitors, etc.

The SDN controller can also act as information extraction mechanism. Furthermore, data can be gained by utilising protocols and tools to retrieve statistics including OpenFlow, sFlow, SNMP, NetFlow, tcpdump, etc. The extracted data or information are made available to the data distribution element.

The data distribution is responsible for carrying data or information from the point of extraction to the analytics, functioning as an interface between them. A different instance of the mechanism described here can also act as the knowledge dissemination element of the Seer platform. The data distribution can either retain information in a persistent manner, e.g. distributed database system, or in a more temporal way, e.g. distributed messaging system. The data distribution should be scalable, elastic, fault tolerant and able to deliver high volume of information with low latency in real time. Examples of distributed messaging systems include Apache Kafka [16], RabbitMQ, AMQP, ActiveMQ, ZeroMQ, etc. Distributed database systems can be NoSQL databases like key-value databases, Redis, Memcached, etc., column store databases like Apache Cassandra [31], Apache HBase, Amazon DynamoDB, etc., document databases as MongoDB, CouchDB, etc. or graph databases like Neo4j. The knowledge dissemination component can also be accompanied by a web service in order to interface with other stake holders of the KCN system.

The data analytics tool is the core of the Seer platform and accommodates its main functionality. It performs data cleaning, annotation and representation as well as modelling, analysis and interpretation. This allows generation of the required knowledge for system prediction and resource optimisation. Input to analytics is also provided by other stakeholders with a form of knowledge. For the platform to achieve continuous delivery, the data analytics should meet the scalability challenges and maintain fault tolerant operation. Online processing should be either provided in a streaming or micro-batching manner, depending on the latency requirements. Exceptional examples of tools include Apache Spark [13], Apache Storm, Apache Flink, Heron [14] etc. The output is forwarded by the knowledge dissemination component to either the knowledge driven control of the same administrative domain, or to other domain's dissemination component.

The component of the platform that utilises the produced knowledge is the so-called knowledge driven control. The SDN controller consumes the knowledge in the logically centralised control plane in order to provide distributed control of the network resources. Computational orchestrators can also utilise the knowledge produced in order to rework storage and compute allocation. Proactive allocation of resources (flows, caches, compute nodes), can result in low latency and highly responsive systems. Overall, easier programmability results in faster adaptation of systems in order to meet the challenges of future systems. Examples of SDN controllers include OpenDaylight [8] and ONOS [9], as mentioned, whereas an example of cloud orchestrator is OpenStack [22].

### B. Prototype

In order to examine the design of Seer, we developed our first prototype focusing on scalability and high availability. We based our implementation on existing open-source tools designed to address requirements of software defined networks and big data analytics. Since the platform inherits the qualities from the comprised components, the core elements had to be well proven, scalable and fault tolerant systems. Therefore, the information extraction element was facilitated by the ONOS controller. An ONOS application was developed in Java to capture events and gather statistics required by the analytics. The ONOS application was publishing the extracted information to Apache Kafka distributed messaging system, which played the role of the data distribution element. On the other end of Kafka, an application running on Apache Spark subscribed to relevant messaging topic and processed the arriving information in micro-batches of one second. Spark played the role of data analytics tool. The derived model was being stored by Spark into a distributed cluster of Apache Cassandra databases. A web micro-service, implemented in Scala and Play framework, was running alongside Cassandra in order to provide a RESTful API for the knowledge dissemination. The output of analytics was also accessible by the ONOS application in an asynchronous manner. The first prototype of Seer is illustrated in Figure 2, while Figure 3 presents the details of Spark configuration.

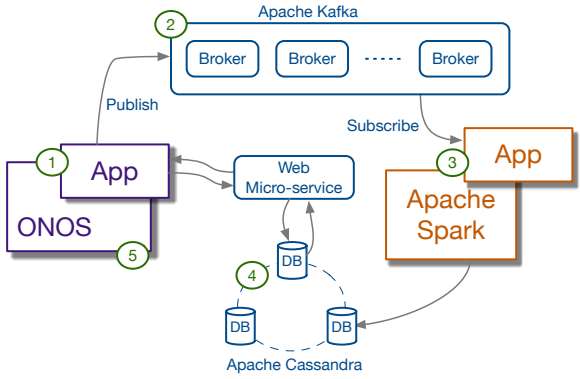

Figure 2. First prototype of the Seer platform composed by ONOS controller, Apache Kafka, Apache Spark, Apache Cassandra and a web micro service attached to Cassandra

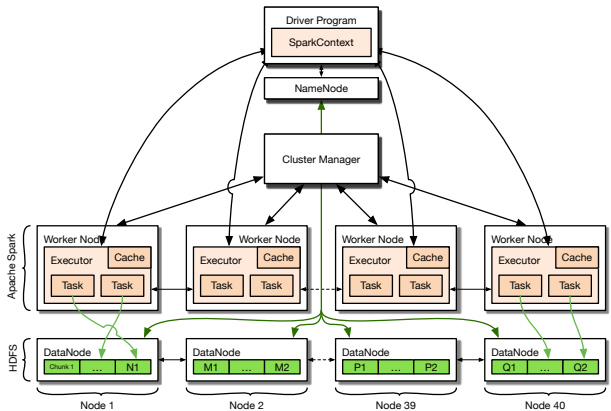

Figure 3. Details of the Apache Spark cluster. The cluster can scale horizontally up to 40 nodes in our lab

The key feature of Seer's first prototype is its modular architecture that facilitates scale-out performance and fault tolerance. All the systems that were used are able of scaling up in order to meet demand. As the data required for analytics grow in volume, the platform can scale up by adding additional instances per element. ONOS can form a cluster comprised of one or more ONOS instances, or nodes, sharing network state with each other. Kafka can be configured in order to form a multi-broker cluster, replicate data and distribute work load. Similarly, Spark forms a master-slave cluster which can scale horizontally and live. Finally, Cassandra can achieve linear scalability without compromising performance while replicating across many nodes. Since every module of Seer can provide functional replication and distribution, the overall platform can achieve high availability.

## V. EVALUATION

To evaluate Seer, we aimed to address the requirements of a fog computing scenario in a smart city environment. We chose to implement a mobility management application that extracts handover information from the network and predicts citizens' movement. The understanding of mobility patterns in a city is valuable knowledge that can drive many edge network applications. The latter include content delivery services, device-to-device communications, autonomous vehicles, augmented reality application and many other that require low latency services, achieved by proactive resource allocation.

For the simulation of the mobility in a smart city, a Python application was created that, given the city's coordinates, acquires the largest employers of the city (universities, hospitals etc.), food shops, health related spots (gyms, sport clubs, etc.) and other recreational corners (pubs, restaurants, etc.) using Google Places API [32]. The simulation builds a 2D probability density function to represent highly commercial zones using kernel functions, as shown in Figure 4. It subsequently generates a range of unique citizens with random locations of personal interest (home, job, gym, club location) and simulates their movement throughout the week. When a citizen is passing close to a Wi-Fi access point (AP) of a given system (i.e. eduroam [33], Bristol, UK), emulated by Mininet [34], the simulation generates a connection between the user's device and the access point. Without loss of generality, we assume that every AP is connected to a different SDN enabled switch. The Wi-Fi system is controlled by an ONOS which hosts an application developed for Seer.

The application inside the ONOS controller subscribes to the host service and captures the handover events which occur when a citizen is switching cells. In order to ensure confidentiality, the ONOS application anonymises the information related to handover before any transmission. The anonymisation process involved the hashing of sensitive information and the frequent (every 24h) re-salting of the used hashes. Every event is published to Kafka following a certain format convention. For our mobility management application, the adopted format was *(id, from, to, timestamp)* which is a 4-tuple of string values. The *id* represents the anonymised MAC address of citizen's mobile device, *from*, and *to*, show the BSSIDs of the APs that the device disconnected from and connected to, respectively. In case the device joins or leaves the Wi-Fi system the fields *from* and *to* are assigned to predefined *null* value. The *timestamp* field features the time and day of the week that the handover event occurred. Using the KCN terminology, the event tuple represents a D1 knowlet.

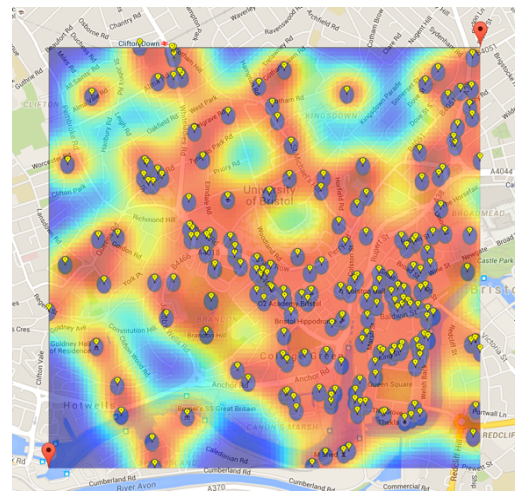

Figure 4. Probability density function representing the commercial zones for mobility simulation in a smart city environment

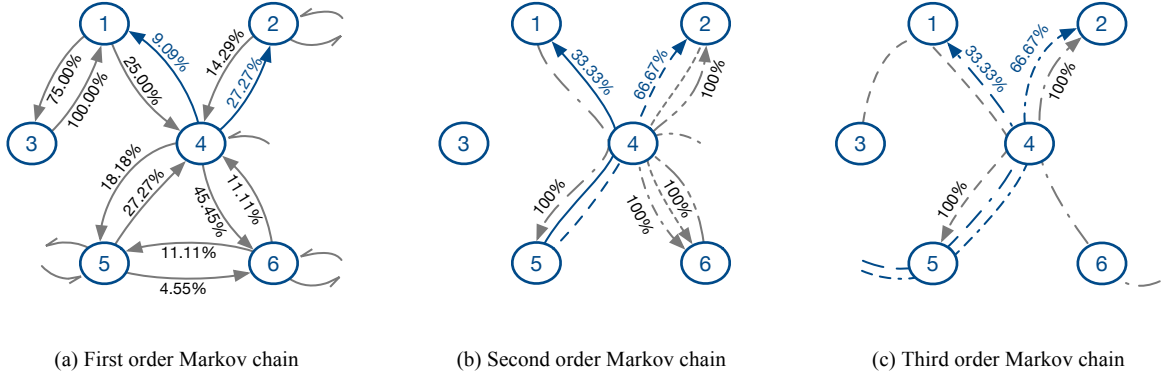

Figure 5. Markov chains describing the transition probability between APs. The higher the order of the chain, the more accurate the prediction and more buffering is required by the algorithm

The Spark streaming application subscribes to the Kafka topic, in which ONOS application publishes events, and retrieves the information as arrives. In order to filter out handovers with large gap between leaving and re-joining timestamps, the application sessionises (i.e. groups together) events that occur in less than $t_{gap}$ time gap between them. So $t_{gap}$ defines the time difference as threshold under which two events are considered part of the same group. For this experiment, $t_{gap}$ was frivolously configured to 300s, due to geographical expansion of the Wi-Fi system. After sessionising per device, the sequence of events is simplified by removing the intermediate events of trivial cases, e.g. devices leaving the Wi-Fi system and then re-joining. The list of sequences is further simplified by excluding sequences of events for devices joining and then leaving immediately without returning back in $t_{gap}$. The *id* information is then removed from every session. Up to this point the format of all sequences is a list of 3-tuples containing *(from, to, timestamp)* for every event in the session.

In order to produce multiple order Markov chains, every session is transformed to *N* variations to represent a Markov Chain of order 1 to *N*, by creating the history of *N*-1 handovers before the last for every handover. For example, the order three variation of a session contains list of events with the following 5-tuple format: *($h_{t-2}$, $h_{t-1}$, from, to, timestamp)*. The $h_{t-2}$ and $h_{t-1}$ represent the BSSIDs of the APs that the device of the session visited before $h_{t-1}$ and *from*, respectively. The Markov chains that represent the knowledge of mobility patterns of the city as realised by eduroam, described as D2 knowlet, are finally stored into Cassandra database.

The web server, which is attached to Cassandra, receives RESTful requests about the probability of movement from a specific AP, given *M* last handover. The server responds to these requests providing the requested knowledge to either external service (KaaS) or ONOS mobility application (KaaT). The prediction of user's movement can subsequently be used to pre-allocate resources, i.e. flow, to switches where the user is predicted to connect next, with the aim to minimise the latency experienced during horizontal handover.

The mobility knowledge is provided in *N* variations, equal to the number of order chain calculated. The service that is requesting the mobility knowledge can choose one of the *N* variations depending on its capabilities and the targeted precision of prediction. Figure 5 shows that the transition from AP 4 to AP 1 is predicted with higher probability in the second and third order. Nevertheless, the increase in precision is not guaranteed.

VI. CONCLUSION AND FUTURE WORK

In this paper we presented Seer, a flexible, highly configurable platform for network intelligence based on SDN, KCN and Big Data principles. Furthermore, we displayed the goal of Seer to accommodate the development of future algorithms and application that target network analytics. By focusing on reliability, the platform aspires to provide scalable, fault tolerant and real-time platform, of production quality. Moreover, Seer intends to leverage the knowledge exchange following the KCN paradigm.

As part of our future work, we plan to evaluate Seer in a Smart City environment and test the mobility application as part of a service chain (see Figure 6). We also plan to extend our first prototype with real-time streaming engines like Apache Storm, Apache Flink or Heron and identify strengths and weaknesses when used in Seer. We hope to reinforce the security characteristics of Seer in order to achieve confidentiality and integrity of sensitive information. Finally, our aim is to provide a holistic solution that will deploy and scale up every module of Seer when required.

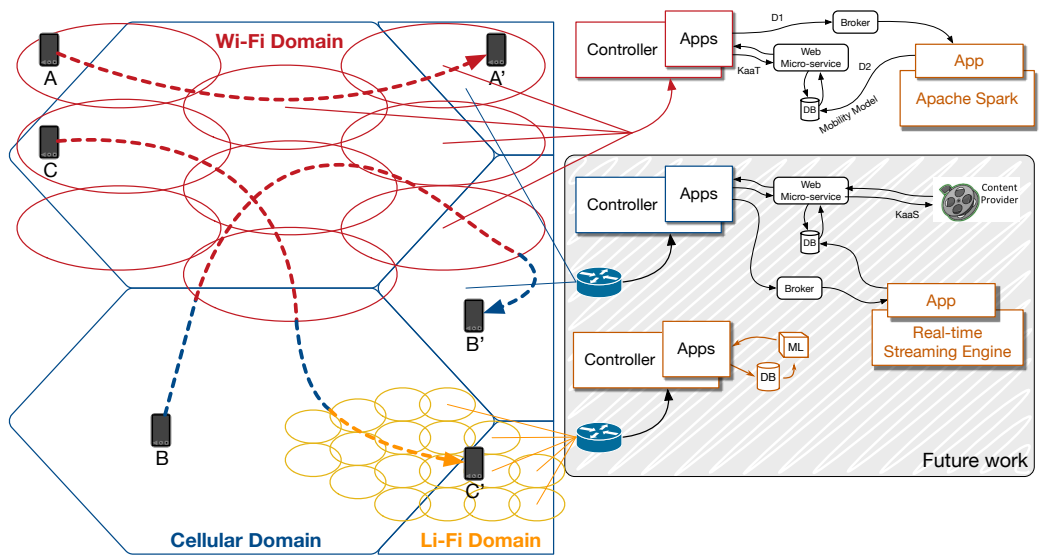

Figure 6. Experimentation of wireless nodes' mobility using simulation in a smart city environment. Stripped area illustrates future work on multiple ADs to utilise KaaS

ACKNOWLEDGEMENT

Kyriakos Sideris would like to thank the following researchers for reviewing this work: Sam Gunner, Paul Anthony Haigh, Yanni Ou, Antonis Papaioannou.